\documentclass{article}
\usepackage[utf8]{inputenc}
\usepackage{authblk}
\usepackage{graphicx}
\usepackage{amsmath}
\usepackage{amssymb}
\usepackage{verbatim}
\usepackage{amsthm,amsfonts,bm}
\usepackage{MnSymbol}
\usepackage{cite}
\usepackage{graphicx}
\usepackage{xcolor}
\usepackage{color}
\usepackage[colorlinks=true, allcolors=blue]{hyperref}
\usepackage{dsfont}
\usepackage{pgf,tikz}
\usetikzlibrary{arrows}
\usepackage{rotfloat}
\usepackage{float}
\usepackage{mathrsfs}
\usepackage{eufrak}
\usepackage[normalem]{ulem}
\usepackage{xcolor,cancel}

\definecolor{qqqqff}{rgb}{0,0,1}
\definecolor{Red}{rgb}{0.9,0,0}
\newcommand{\p}{\partial}

\title{5-Dimensional $SO(1,4)$-Invariant Action as an Origin to the Magueijo-Smolin Doubly Special Relativy Proposal}

\author{B. F. Rizzuti}
\author{G. F. Vasconcelos Jr.}

\affil{Departamento de F\'isica, Universidade Federal de Juiz de Fora, MG, 36.036-900, Brazil}

\date{}                     %% if you don't need date to appear
\begin{document}

\maketitle

\begin{abstract}
In this paper we discuss how the Magueijo-Smolin Doubly Special Relativity proposal may be obtained from a singular Lagrangian action. The deformed energy-momentum dispersion relation rises as a particular gauge, whose covariance imposes the non-linear Lorentz group action. Moreover, the additional invariant scale is present from the beginning as a coupling constant to a gauge auxiliary variable. The geometrical meaning of the gauge fixing procedure and its connection to the free relativistic particle are also described. 
\end{abstract}
\thispagestyle{empty}
\newpage
\section{Introduction}
\label{sec.intro}

The idea to introduce another invariant scale into relativistic regimes is not new. In \cite{snyder_quantized_1947}, the author presents a Lorentz invariant quantized spacetime. It was an attempt to avoid divergences in field theories. More recently, quantum gravity (QG) calculations seem to reinforce the existence of a fundamental scale, implying discreteness of areas and volumes \cite{rovelli_discreteness_1995}. In particular, in \cite{garay_quantum_1995} different approaches to quantum gravity indicate a lower bound to distance measurements. This convergence would confirm that at small scale distances, the very spacetime structure would be discrete. This is not the only issue to be tackled throughout the road to a consistent QG theory though. When gravity is taken into account spacetime becomes alive as a dynamical quantity. Can we try to insert this allegedly new scale in a partial regime where gravity/curvature effects can be neglected? The answer gives rise to the so-called Doubly Special Relativity (DSR) models \cite{amelino-camelia_relativity_2002, magueijo_lorentz_2002}. They have first appeared a couple of years after the enlightening aforementioned paper \cite{garay_quantum_1995}. With different motivations, such as cosmological \cite{amelino-camelia_planck-scale_2001}, physical \cite{GORJI2017113} or even mathematical \cite{deriglazov_position_2005}, the central concept of these trials was the same: to insert another invariant scale, besides the speed of light, while keeping an intermediate regime with fixed spacetime as background. A list of facts and myths about the DSR proposals may be found in \cite{amelino-camelia_doubly-special_2010}.

One of the most celebrated DSR models was proposed by J. Magueijo and L. Smolin \cite{magueijo_lorentz_2002}, which we call MS for short. One of its issues was the lack of a fully covariant spacetime description, since it was set from the beginning on the energy-momentum space. Actually, the construction of DSR Lagrangian models starting from a four-dimensional spacetime is a rather delicate issue, see, for example, \cite{rizzuti_comment_2010}. MS have specified a modified dispersion relation,
\begin{equation}
    p_\mu p^\mu = m^2c^2(1 + \xi p^0)^2,\label{dispersion_MS}
\end{equation}
accompanied by the non-linear representation of the Lorentz group
\begin{equation}
    p'^{\mu} = \frac{\Lambda^\mu_{\;\nu} p^\nu}{1 + \xi(p^0 - \Lambda^0_{\;\nu} p^\nu)},\label{momentum_lorentz}
\end{equation}
which keeps \eqref{dispersion_MS} unaltered. The variable $\xi$ stands for the invariant scale, related to the Planck length. In fact, $p^\mu = (-\frac{1}{\xi},0,0,0)$ remains intact under \eqref{momentum_lorentz}. A possible experimental confirmation to such Lorentz violation scenario would give a possible fingerprint of QG effects, as explained previously. 

This work is intended to show how the MS kinematic rules derived from \eqref{dispersion_MS} and \eqref{momentum_lorentz} can be deduced from a singular particle model. This path has already been taken in a previous work \cite{rizzuti_five-dimensional_2011}. For the latter, the invariant scale $\xi$ enters into the game by an \textit{ad hoc} gauge fixation. This rather non-trivial choice is made only to produce the expected dispersion relation \eqref{dispersion_MS}. In this work a slightly different Lagrangian from the one in \cite{rizzuti_five-dimensional_2011} will be presented. The model lives in a flat 5-dimensional spacetime whereas the corresponding phase space 
is constrained to a constant curvature hypersurface. In the current case, the parameter $\xi$ is present since the beginning. Moreover, we present a clear meaning for it. In effect, it is nothing but the coupling constant that connects a gauge auxiliary variable to the dynamical sector of the action. 

The hunch of considering 5-dimensional  models to the DSR arena is not new. It may be traced back to the papers \cite{linear_5d, mignemi_5d}. The authors in \cite{linear_5d} have shown that the MS transformation is fraction-linear in a 4-dimensional (real) projective space, leading to the corresponding linear action in five dimensions. Whereas \cite{mignemi_5d} is based upon the identification of the de Sitter space with the space of momenta in the DSR context. Actually, the de Sitter space has also been widely considered in Doubly Special Relativity regimes: its radius provides an additional natural candidate for observer-independent scales \cite{jkg, smignemi}. We shall not consider it here for two main reasons. In the configuration sector, it would lead to a curved background and we are interested in an intermediate regime with flat (and fixed) space-time. On the other hand, our space of momenta is parametrized by a cone, see the constraint \eqref{nonsitter} in advance. Hence it prevents our model from being described by a de Sitter space. 

Another possible alternative for DSR models is to look for the conformal group. For example, in \cite{alexei_conformal}, the conformal vector, which parametrizes the transformation, defines the new invariant scale. While the model has a nontrivial space-time metric, the corresponding four-dimensional scalar curvature vanishes. It presents, though, a curved 3-dimensional space-like slice. As explained before, at this stage we would like to avoid any curvature/gravitational effect whatsoever.

The paper is divided as follows. In Section \ref{sec.2} we present a singular Lagrangian action with global $SO(1,4)$ invariance as well as a local symmetry. This model is built on a five-dimensional position space and, as a consequence of its singularity, is endowed with a set of first class constraints, intrinsically connected to the gauge symmetries \cite{deriglazov_generalization_2011}. In Section \ref{sec.3} we apply Dirac's hamiltonization procedure for singular theories \cite{dirac_lectures_2001, deriglazov_classical_2017} and obtain the Hamiltonian equations of motion, together with the explicit first class constraints of the system. Section \ref{sec.4} is designed to shed some light on the physical sector of the model. In Section \ref{sec.5} we discuss how the free relativistic particle and the Magueijo-Smolin proposal can be related by specific gauge choices. Moreover, in Section \ref{sec.6} we propose geometrical interpretations of the previous results and Section \ref{sec.7} is left for the conclusions.

\section{The 5-dimensional Lagrangian Model}
\label{sec.2}

It is a standard procedure to construct mechanical or field models with more variables than the physical ones present on them. This way, a fully global linear Lorentz group covariance can be guaranteed. Consequently, the price to be paid is that the models carry constraints between degrees of freedom to assure that not all among the variables are observables. We point out that the action of the group of covariance acts in a slightly non-linear way upon the physical sector of the models. Standard examples of such models are the free relativistic particle \cite{rizzuti_square_2019} and the electrodynamics Lagrangian action \cite{deriglazov_generalization_2011}.

For the MS DSR, as exposed through \eqref{momentum_lorentz}, one starts with a non-linear realization of the Lorentz group on the four dimensional space of energy-momentum, parametrized by $\{p^\mu\}$. Thus, our suggestion here is to start with a five dimensional singular Lagrangian model with a set of first class constraints. As it will become clear in a while, this Lagrangian has global $SO(1,4)-$invariance. The non-linear $SO(1,3)-$action will be achieved by slicing the phase space through the gauge fixing, for the (first class) constraints. Throughout the paper, unless stated otherwise, dots over quantities mean derivatives with respect to $\tau$, that is, $\dot{\gamma} := \frac{d\gamma}{d\tau}$. The construction proceeds as follows.

Consider the five dimensional configuration space para- metrized by $\{x^A,g\}$, where the indices $A$, $B$, $C$, $...$ take the values 0, 1, 2, 3, 5, and $g$ is an auxiliary variable. Our particle model is described by the action
\begin{equation}
    S = \int d\tau \left[\frac{m}{2}\eta_{AB}\mathcal{D}x^A\mathcal{D}x^B - \xi g\right] := \int d\tau L.\label{main_action}
\end{equation}
Here $\tau$ is an (arbitrary) evolution parameter and $m$ shall be interpreted as the rest mass of the particle. We also point out that $\mathcal{D} := \frac{d}{d\tau} - g$ can be seen as an analogue of a covariant derivative to gauge theories, with gauge field $g$. This fact will be confirmed soon. However, we firstly note that the factor $\xi$ in second term in \eqref{main_action} is interpreted as a coupling constant that connects the gauge field $g$ with the dynamical sector of the model. The configuration space is endowed with the (pseudo)-metric $\eta_{AB} = (+1,-1,-1,-1,-1)$. Additionally, it is clear that the action $\eqref{main_action}$ has global $SO(1,4)$-invariance,
\begin{equation}
    x^A \rightarrow x'^{A} = \Lambda^A_{\;\;B}x^B;\;\;\;\forall \Lambda \in SO(1,4).
\end{equation}
Besides the global symmetry, the action is also invariant under the local exact transformations
\begin{align}
    \tau \to \tau'(\tau) = \Gamma(\tau);\;\;\;\;\; \frac{d\tau'}{d\tau} = \gamma^2(\tau),\label{transfo_tau}\\
    x^A(\tau) \to x'^{A}(\tau') = \gamma(\tau)x^A(\tau),\label{transfo_xA}\\
    g(\tau) \to g'(\tau') = \frac{\dot{\gamma}(\tau)}{\gamma^3(\tau)} + \frac{g(\tau)}{\gamma^2(\tau)},\label{transfo_g}
\end{align}
parametrized by the arbitrary function $\gamma$. In effect, we first note that the transformation of the derivative $\mathcal{D}x^A$ under these is
\begin{equation}
    \mathcal{D}x^A \to \mathcal{D}'x'^{A} = \frac{1}{\gamma}\mathcal{D}x^A.\label{claim_covariant_derivative}
\end{equation}
In turn, this calculation and the simple result \eqref{claim_covariant_derivative} justify our claim that $\mathcal{D}$ may be identified as a covariant derivative, with $g$ being the corresponding gauge field. This is the very structure one may find in the case of electrodynamics and its interaction with the electron field, when the global $U(1)$ transformation is graduated to a local one \cite{weinberg_quantum_1996}. Now, if one plugs \eqref{transfo_tau}, \eqref{transfo_xA}, \eqref{transfo_g} and \eqref{claim_covariant_derivative} back on the action \eqref{main_action} then the result is
\begin{equation}
   S \to S' =  \int L'd\tau' = \int d\tau \left[L + \frac{d}{d\tau}\left(-\xi ln \gamma\right)\right].
\end{equation}
Since $L$ and $L'$ differ by a total derivative term, the transformations \eqref{transfo_tau}, \eqref{transfo_xA} and \eqref{transfo_g} are local symmetries of the action $S$, as stated previously.

The presence of gauge symmetries is linked to the appearance of first class constraints through the Dirac recipe for singular systems \cite{deriglazov_generalization_2011}. This is the standard procedure to investigate singular mechanical models and shall be done in the next section.

\section{Hamiltonian formulation and constraints}
\label{sec.3}

In this section we will construct the Hamiltonian version of the action \eqref{main_action}. This is a singular model since the Hessian matrix, whose components are,
\begin{equation}
    \frac{\p^2L}{\p\dot{Y}^a\p\dot{Y}^b};\;\;\;\;Y^a=\{x^A,g\},
\end{equation}
has a null determinant and Dirac's hamiltonization prescription can be used. Among the advantages of using the Hamiltonian over the Lagrangian formulation, the main one is to uncover constraints between degrees of freedom as part of the Hamiltonian equations.

The first step in the prescription consists of defining the conjugate momenta: $p_a := \frac{\p L}{\p\dot{Y}^a}$. With more details,
\begin{align}
    p_A &= \frac{\p L}{\p\dot{x}^A} = m\eta_{AB}(\dot{x}^B - gx^B),\label{momenta_pA}\\
    p_g &= \frac{\p L}{\p\dot{g}} = 0.\label{momenta_g}
\end{align}
They are used as algebraic equations to obtain the velocities in terms of configuration and momenta variables. From Eq. \eqref{momenta_pA} we have
\begin{equation}
    \dot{x}^A = \eta^{AB}\left(\frac{p_B}{m} + gx_B\right),
\end{equation}
where $\eta^{AB} (\eta_{CD})$ may be used to raise (lower) indices. Eq. \eqref{momenta_g}, in turn, is a primary constraint. With these quantities in hand we are now able to write the Hamiltonian of the system,
\begin{align}
    H(Y^a,p_a) &= \left(p_a\dot{Y}^a - L\right)|_{\eqref{momenta_pA},\eqref{momenta_g}} + v_gp_g \nonumber\\ 
    &= \frac{1}{2m}\eta^{AB}p_Ap_B + g\eta^{AB}p_Ax_B + \xi g + v_gp_g.\label{complete_hamiltonian}
\end{align}
It is defined on the extended phase space parametrized by $\{Y^a,p_a,v_g\}$, where $v_g$ is the Lagrange multiplier for the constraint $T_1 := p_g = 0$. The Poisson brackets are defined canonically,
\begin{equation}
    \{.,.\} = \frac{\p}{\p Y^a}\frac{\p}{\p p_a} - \frac{\p}{\p p_a}\frac{\p}{\p Y^a},
\end{equation}
allowing us to write the equations of motion
\begin{align}
    \dot{x}^A = \{x^A,H\} &= \frac{1}{m}\eta^{AB}p_B + gx^A;\\
    \dot{p}_A = \{p_A,H\} &= -gp_A;\\
    \dot{g} = \{g,H\} &= v_g;\\
    T_1 = p_g &= 0.
\end{align}

Due to a consistency condition, it is expected that $T_1$ remains equal to zero throughout the passage of time. Henceforth, one finds the equations of the following stages of the Dirac procedure as algebraic consequences of this system. In effect,
\begin{equation}
    0= \{T_1,H\} \Rightarrow T_2 := p_Ax^A + \xi = 0.
\end{equation}
By applying the same line of reasoning one finds a third-stage constraint,
\begin{equation}\label{nonsitter}
    0= \{T_2,H\} \Rightarrow T_3 := \eta^{AB}p_Ap_B = 0.
\end{equation}
The procedure stops at this step, since the time evolution of $T_3$ does not bring any new information. Therefore, the complete set of constraints of this model is $\{T_1,T_2,T_3\}$. Moreover, they are all first class constraints. Indeed, we have
\begin{align}
    \{T_1,T_2\} = \{T_1,T_3\} = 0;\\
    \{T_2,T_3\} = 2T_3. \quad
\end{align}

\section{Physical Sector}
\label{sec.4}

We begin by pointing out that throughout this section and the rest of the paper the Greek letters $\mu,\nu,...$ shall take values in $\{0,1,2,3\}$, whilst the letters $i,j,k,...$ shall take values in $\{1,2,3\}$, unless stated otherwise. Our model is described by a singular Lagrangian, subject to the local symmetries \eqref{transfo_tau}, \eqref{transfo_xA} and \eqref{transfo_g}. It implies that all our initial variables have ambiguous evolution and, thus, are not suitable candidates to be observables. The lack of a single solution for the equations of motion is exposed through the Dirac procedure. In fact, the Lagrange multiplier $v_g$ in \eqref{complete_hamiltonian} cannot be found as an algebraic consequence of the equations of motion. Thus, $g$ enters into the game as an arbitrary function of the evolution parameter $\tau$. The same train of though applies to both $x^A$ and $p_B$, since the corresponding equations of motion have an explicit dependence on $g$. In order to skirt this issue we separate the fifth dimension, defining the variables
\begin{align}
    z^\mu = \frac{x^\mu}{x^5},\\
    \pi_\mu = \frac{p_\mu}{p_5},
\end{align}
These quantities are not chosen randomly. We simply follow the same steps taken on the standard analysis of the free relativistic particle model \cite{rizzuti_square_2019}, or, by the same token, the semiclassical spinning particle model \cite{alexei_rizzuti_genaro_2012}, in order to remove the so discussed arbitrariness our model possesses. First of all, we point out that these quantities remain unaltered under the local symmetries,
\begin{align}
    z^\mu &\to z'^\mu = \frac{x'^\mu}{x'^5} = z^\mu,\label{observable_candidate_1}\\
    \pi_\mu &\to \pi'_\mu = \frac{p'_\mu}{p'_5} = \pi_\mu\label{observable_candidate_2}.
\end{align}
Therefore, they are suitable candidates for observables of our model. In particular, we may promptly write the associated equations of motion
\begin{align}
    \dot{z}^\mu &= \frac{1}{m}\lambda(\pi^\mu - z^\mu),\label{z_dot}\\
    \dot{\pi}_\mu &= 0,\label{pi_dot}
\end{align}
where $\lambda := \frac{p_5}{x^5}$. Evidently, the equations of motion for the $(z,\pi)$-sector resemble those of the free relativistic particle. The ambiguity due to the arbitrary parameter $\lambda$ is related to the reparametrization invariance of the theory. Henceforth, $z^\mu(\tau)$ can be interpreted as the parametric equations of the physical variables $z^i(z^0)$. The latter may be obtained by inverting the expression $z^0=z^0(\tau) \leftrightarrow \tau = \tau(z^0)$ and substituting it back in $z^i(\tau(z^0)) \equiv z^i(z^0)$.  

%\sout{If we set, as usual, $z^0 = ct$ and consider $c$ as the speed of light, we get}
%\begin{equation}
%    \cancel{\frac{dz^i}{dz^0} = \frac{\dot{z}^i}{\dot{z}^0} = \frac{\pi^i - z^i}{\pi^0 - z^0} \Rightarrow \frac{dz^i}{dt} = c\frac{\pi^i - z^i}{\pi^0 - z^0}.}
%\end{equation} 
%
%\sout{In terms of the $\pi_\mu$ variables, the constraint $\eta^{AB}p_Ap_B = 0$ reads}
%\begin{equation}
%\cancel{
%   \eta^{\mu\nu}p_\mu p_\nu - (p_5)^2 = 0 \Rightarrow \eta^{\mu\nu}\pi_\mu\pi_\nu = 1.} \label{31}
%\end{equation}
%\sout{Finally, we may resolve $\pi_0$ from \eqref{31},} 
%\begin{equation}
%  \cancel{\pi_0 = \sqrt{1 + \delta^{ij}\pi_i\pi_j}} 
%\end{equation}
%\sout{where $\delta^{ij}$ is the Kronecker delta. Hence, in the limit $|\pi_i| \rightarrow +\infty$, we obtain}
%\begin{equation}
%    \cancel{\frac{dz^i}{dt} = c\frac{\pi^i - z^i}{\sqrt{1 + \delta^{ij}\pi_i\pi_j} - z^0} \rightarrow c\hat{\pi}^i,}
%\end{equation}
%\sout{where $\hat{\pi}^i := \frac{\pi^i}{|\pi^i|}$. This result reveals that the physical variables in our system describe a free particle which, particularly, is moving in a straight line with a speed bounded above by $c$.}

We may solve the system of differential equations formed by \eqref{z_dot} and \eqref{pi_dot}. As it was discussed previously, $\lambda$ is an arbitrary parameter and thus we choose $\lambda = m$. Naturally, the solutions of $\eqref{pi_dot}$ are constants with respect to $\tau$, that is, $\pi_\mu(\tau) = const..$ On the other hand, solutions of \eqref{z_dot} are given by
\begin{equation}
    z^\mu(\tau) = \pi^\mu + Z^\mu e^{-\tau},
\end{equation}
where the quantities $Z^\mu$ arise as integration constants. By declaring $z^0(\tau)$ as the evolution parameter of the system, we get
\begin{equation}
	z^0(\tau) = \pi^0 + Z^0e^{-\tau} \Rightarrow e^{-\tau} = \frac{z^0 - \pi^0}{Z^0}.
\end{equation}
In turn, this expression is the implicit representation of $\tau = \tau(z^0)$. Therefore,
\begin{equation}
	z^i(\tau(z^0)) = z^i(z^0) = \pi^i + Z^i\left(\frac{z^0 - \pi^0}{Z^0}\right) = v^iz^0 + (\pi^i - v^i\pi^0),\label{sol_freepar_DSR}
\end{equation}
where $v^i := \frac{Z^i}{Z^0}$. It follows that $\frac{d^2z^i}{d{z^0}^2} = 0$, \textit{i.e.}, the physical sector describes a free particle. This is an interesting result, because the original action \eqref{main_action} is not invariant under translations. Moreover, \eqref{sol_freepar_DSR} was obtained regardless of the fixation of the gauge $g=0$, as we shall do in a while. We point out, though, that the constants $v^i$ are not bounded above by the speed of light. Thus the particle is not relativistic.

To conclude this section, we point out that our model bears the same number of degrees of freedom of \emph{DSR} particles. In fact, our extended phase space is parametrized by $\{x^A,p_B,g,p_g\}$, counting a total of 12 degrees of freedom. Taking into account that each first class constraint rules out 2 spurious degrees of freedom (after gauge fixation), we are left with $12 - 2\times 3 = 6$ physical degrees of freedom, as expected. 3 of those are related to the position of the particle as a function of time, $x^i(x^0)$, and the 3 remaining ones come from the deformed dispersion relation \eqref{dispersion_MS}. Needles to say, the same arguments hold true for the free relativistic particle.

\section{Gauge Fixing and MS DSR proposal}
\label{sec.5}

Our next task consists of slicing out the momentum sector of the phase space. This means that particular gauges will be fixed for our first class constraints in the form of hyperplanes. In turn, they will reproduce both the $(i)$ free relativistic particle (FRP) and $(ii)$ the Magueijo-Smolin DSR proposal. The relation between these two models shall be discussed in the next section.

Furthermore, as it is usual in gauge theories, the local symmetries are not preserved after the gauges are fixed. Nonetheless, one can search for their combinations that retain the gauge condition. Following this prescription, both cases $(i)$ and $(ii)$ may be derived.

\subsection{Free Relativistic Particle Gauge}
\label{subsec.frp}

To reproduce the FRP dynamics we fix the following gauges for the constraints $T_1 = p_g$ and $T_2 = p_Ax^A + \xi$, forming second class pairs of constraints:
\begin{align}
    g &= 0;\;\;\;\;\;\;\{g,p_g\} = 1,\\
    p^5 &= mc;\;\;\;\{p^5 - mc,p_Ax^A + \xi\} = mc.\label{gauge_plane_p5_MS_DSR}
\end{align}
Moreover, we impose invariance (both global and local) for the gauge conditions. First, for the $g-$sector, we have
\begin{equation}
    g = 0 \Leftrightarrow g' = \frac{\dot{\gamma}(\tau)}{\gamma^3(\tau)} + \frac{g(\tau)}{\gamma^2(\tau)} = 0 \Rightarrow \gamma = const.
\end{equation}
Under local and global transformations, the momenta $p^A = m\mathcal{D} x^A$ transform as
\begin{equation}
    p^A \to p'^A = \frac{1}{\gamma}\Lambda^A_{\;\;B}p^B,\;\;\;\Lambda^A_{\;\;B}\in SO(1,4).
\end{equation}
We restrict ourselves to the subgroup of  $SO(1,4)$ whose elements are $\Lambda^\mu_{\;\;\nu} \in SO(1,3)$, such that
\begin{equation}
\Lambda^A_{\;\;B} = \left(\begin{array}{@{}ccccc|c@{}}
  {} & {} & {} & {} & {} & 0 \\
  {} & {} & { \Lambda^{\mu}{}_{\nu}} & {} & {} & 0 \\
  {} & {} & {} & {} & {} & 0 \\ 
  {} & {} & {} & {} & {} & 0 \\ \hline
  {} &  0 & 0 & 0 & 0 & 1 
\end{array}\right)
\end{equation}
%\begin{equation}
%\Lambda^A_{\;\;B} = \begin{pmatrix}
%\begin{matrix}
%\Lambda^\mu_{\;\;\nu}
%\end{matrix} &
%\begin{matrix}
%0\\
%0\\
%0\\
%0\\
%\end{matrix}\\
%\hline
%\begin{matrix}
%0 & 0 & 0 & 0
%\end{matrix} & \begin{matrix}
%1
%\end{matrix}
%\end{pmatrix},
%\end{equation}
since for $p^5 = mc$ the remaining sector of $\Lambda^A_{\;\;B}$ induces boosts in the fifth dimension, which can be seen as translations in four dimensions, that is,
\begin{equation}
    p'^\mu = \Lambda^\mu_{\;\;A}p^A = \Lambda^\mu_{\;\;\nu}p^\nu + \mathcal{P}^\mu.
\end{equation}
Here $\mathcal{P}^\mu = \Lambda^\mu_{\;\;5}mc = const.$. We follow the same steps for the $p^A$-sector,
\begin{equation}
    p^5 = mc \Leftrightarrow p'^5 = mc = \frac{1}{\gamma}\Lambda^5_{\;\;A}p^A = \frac{1}{\gamma}mc \Rightarrow \gamma = 1,
\end{equation}
which fixes the function $\gamma$.

Finally, we may assemble the results that have been obtained previously. The dynamics reads
\begin{equation}
    \dot{x}^5 = c,\;\;\;\;\;\;\dot{x}^\mu = \frac{1}{m}p^\mu\;\;\;\;\;\;\mathrm{and}\;\;\;\;\;\;\dot{p}_\mu = 0\label{dynamics}
\end{equation}
The $p_\mu$ coordinates are restricted to the mass-shell relation
\begin{equation}
    \eta^{\mu\nu}p_\mu p_\nu = m^2c^2,\label{mass-shell_relation}
\end{equation}
which was obtained by substituting the gauge \eqref{gauge_plane_p5_MS_DSR} back into the constraint $\eta^{AB}p_A p_B = 0$. With $\gamma = 1$, the momenta $p_\mu$ transform as
\begin{equation}
    p'^\mu = \Lambda^\mu_{\;\;\nu}p^\nu.\label{momenta_transformations}
\end{equation}
The equations \eqref{dynamics}, \eqref{mass-shell_relation} and \eqref{momenta_transformations} allow us to interpret the gauge fixed version as a free relativistic particle, with mass $m$ and momenta $p_\mu$. This gauge may be seen as a subset of solutions of the physical sector, described in Section \ref{sec.4}. In fact, in the latter we have no restriction to the particle's speed encoded by \eqref{mass-shell_relation}. Moreover, the fifth dimension in the configuration space is just the arbitrary evolution parameter, $x^5 = c\tau + const.$.

\subsection{Magueijo-Smolin DSR Gauge}
\label{MS_gauge}

We follow the same steps we have made so far. Once again, we take $g=0$ for the constraint $T_1 = p_g$. However, instead of the hyperplane given by \eqref{gauge_plane_p5_MS_DSR}, we work with a slightly rotated version of it and fix the MS DSR gauge,
\begin{equation}
    p^5 = mc(1 + \xi p^0)\label{MS_DSR_gauge},
\end{equation}
for the constraint $T_2 = p_Ax^A + \xi$, forming a second class pair
\begin{equation}
    \{p^5 - mc\xi p^0 - mc, p_Ax^A + \xi\} = mc \neq 0.
\end{equation}

Invariance of $g$ implies, as before, the fixation of the local parameter $\gamma = const.$. It can be fully determined by imposing the invariance of the MS DSR gauge,
\begin{equation}
    p^5 = mc(1 + \xi p^0) \Leftrightarrow p'^5 = mc(1 + \xi p'^0) \Rightarrow \gamma = 1 + \xi(p^0 - \Lambda^0_{\;\;\mu}p^\mu).
\end{equation}
Thus, we are left with a free particle bounded to the MS DSR kinematical predictions,
\begin{equation}
\dot{x}^5 = c(1 + \xi p^0),\;\;\;\;\;\;\dot{x}^\mu = \frac{1}{m}p^\mu\;\;\;\;\;\;\mathrm{and}\;\;\;\;\;\;\dot{p}_\mu = 0,\label{dynamics_MS_DSR}
\end{equation}
where the momentum is restricted to the deformed dispersion relation
\begin{equation}
    \eta^{\mu\nu}p_\mu p_\nu = m^2c^2(1 + \xi p^0)^2\label{hyperplane_MS_DSR_gauge}
\end{equation}
and transforms accordingly,
\begin{equation}
    p'^\mu = \frac{\Lambda^\mu_{\;\;\nu}p^\nu}{1 + \xi(p^0 - \Lambda^0_{\;\;\mu}p^\mu)}.\label{transformation_momenta_MS_DSR_gauge}
\end{equation}
As claimed, equations \eqref{dynamics_MS_DSR}, \eqref{hyperplane_MS_DSR_gauge} and \eqref{transformation_momenta_MS_DSR_gauge} follow from the non-standard gauge \eqref{MS_DSR_gauge} and represents the MS DSR proposal. Once more, the $x^5$ coordinate is proportional to the arbitrary parameter $\tau$. We highlight that eq. \eqref{dynamics_MS_DSR} does not imply that $x^5$ evolves with a speed faster than the speed of light. $\tau$ can be reparametrized and has no physical interpretation. A change of scale could suppress $\Tilde{c}:=c(1+\xi p^0)$: $\tau \to \tau' = \frac{c}{\Tilde{c}}\tau \Rightarrow x^5 = c\tau + const.$. 

\section{Geometrical meaning for the gauge fixing procedure}
\label{sec.6}

\begin{figure}[!h]
    \centering
    \includegraphics[scale=0.5]{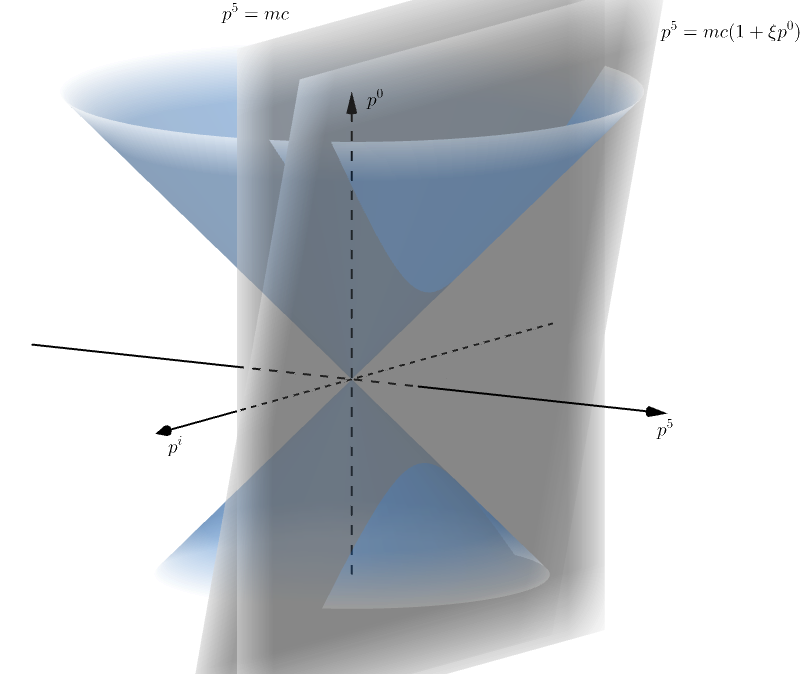}
    \caption{FRP and MS DSR gauges slicing out the hypersurface of momenta.}
    \label{fig:cone}
\end{figure}

It is usual to set DSR models initially on the space of conserved energy - momentum. With this perspective in mind let us analyze the $p_A$-sector of our proposal. To begin with, it is governed by the constraint $T_3 = \eta^{AB}p_A p_B = 0$. If we merge the $p^i$ coordinates among the indices $A,B$, then we can sketch its geometrical representation, which is a 5-dimensional cone. The FRP gauge is nothing but the plane $p^5 = mc$, which, when intersecting the previous structure, results in the standard hyperboloid on the momenta coordinates $\{p_\mu\}$, described by the usual mass-shell condition. This intersection $(\{T_3 = 0\}\cap\{p^5 = mc\})$ is shown in Fig. \ref{fig:cone}. 
On the other hand, the MS DSR gauge corresponds to the hyperplane characterized by $p^5 = mc(1 + \xi p^0)$, which is also exhibited in the same figure.

The angle between the planes that define both gauges is given by
\begin{equation}\label{49}
    \theta := \arctan(mc\xi).
\end{equation}
It is clear from the Fig. \ref{fig:cone} that the rotation of one of the gauges around the $p^i$-sector by this angle generates the other.

Let us now simulate a scenario where DSR effects could be feasible. We consider, for instance, a proton close to the GZK threshold \cite{greisen_end_1966, zatsepin_upper_1966}, $p_{GZK}^0 = \frac{5\times10^{19}eV}{c}$. For this estimate, we assume $\xi = \frac{1}{10^3\times p^0_{GZK}}$, as $\xi p^0$ should be much lesser than 1. Therefore, in this case one finds
\begin{equation}
    \theta \approx 2\times10^{-14}.
\end{equation}
Since $\theta << 1$, our calculation indicates that it would be difficult to detect a DSR effect when comparing it to measurements obtained according to the standard relativistic kinematics predictions.

\section{Conclusions}
\label{sec.7}

In this work we proposed a singular Lagrangian model \eqref{main_action} on a five-dimensional spacetime which, after fixing specific gauges, reproduces both the free relativistic particle and the Magueijo-Smolin doubly special relativistic kinematics. The Lagrangian was chosen so that it would be globally invariant under the $SO(1,4)$ group of symmetries. In turn, the local symmetries of the configuration space variables were presented. Due to the singularity of the system, its Hamiltonian formulation was constructed according to Dirac's hamiltonization procedure. This allowed us to explicitly uncover the full set of first class constraints between degrees of freedom from the Hamiltonian equations of motion.

Since the model presents first class constraints, it was discussed that not all of its variables could be observables of the system. In effect, we discussed that the initial set of variables had ambiguous evolution and were not in the physical sector of the model. However, separation of the fifth spacial dimension from the rest of the variables allowed us to write suitable candidates for observables of the model. Besides that, it was discussed that the equations of motion of these quantities resembled those of the free relativistic particle on flat four-dimensional spacetime. Moreover, it was shown that such particle obeyed the standard relativistic dispersion relation and its speed was bounded above by the speed of light, while bearing the same number of degrees of freedom of DSR particles.

Finally, after fixing particular gauges for the first class constraints both the FRP and the MS DSR dynamics could be reproduced from this model. In turn, the MS gauge implied the model reduced to a free relativistic on the deformed four-dimensional space. Moreover, this approach of slicing out the momentum sector of the phase space with planes led us to a geometrical interpretation of the relation of these two systems: the scale $\xi$ defines the angle between the two planes, according to \eqref{49}.

Although new perspectives indicate that the MS is just a particular type of a broader class of DSR proposals \cite{JAFARI2020135735}, our results show that the MS DSR turns out to be one particular gauge of a FRP, living on a hypersurface of constant curvature. These last two observations may suggest that the rise of new observable physical effects within this context would be demanding to detect.     

\section*{Acknowledgement}

This work is supported by XXIX PIBIC/CNPq/UFJF - 2020/2021, project number ID-47862.
During the revision process of this manuscript, Prof. A. Deriglazov drew our attention to the behavior of the physical sector of our model. Although the initial Lagrangian has no translation invariance, the physical degrees of freedom describe a free particle. This issue is currently under investigation. The authors would like to express their gratitude to the valuable insights.

%\bibliography{biblio.bib}

%\bibliography{mybiblio}{}

%\bibliographystyle{apa}
%\bibliographystyle{ieeetr}
%\clearpage

\end{document}